\documentstyle[aps,preprint,epsf,epsfig]{revtex}

\def\dfrac{\displaystyle\frac}
\def\bolpor{
\kern-.2em\raise.4ex\hbox{$>$} \kern-1.em\lower.4ex\hbox{\small
${\sim}$}
}
\def\menpor{
\kern-.2em\raise.4ex\hbox{$<$} \kern-1.em\lower.4ex\hbox{\small
${\sim}$}
}
\def\i{\imath}
\def\o{\omega}
\def\g{\gamma}

\def\a{\alpha}
\def\o{\omega}
\def\b{\beta}

\def\hp{\hat{p}}

\def\d{\partial}
\def\noi{\noindent}
\def\ve{\varepsilon}

\newcommand{\Eq}[1]{Eq.(\ref{#1})}

\newcommand{\refc}[1]{Ref.~\cite{#1}}
\newcommand{\refs}[1]{Refs.~\cite{#1}}
\newcommand{\bea}{\begin{eqnarray}}
\newcommand{\eea}{\end{eqnarray}}
\newcommand{\be}{\begin{equation}}
\newcommand{\ee}{\end{equation}}
\newcommand{\bc}{\begin{center}}
\newcommand{\ec}{\end{center}}
\newcommand{\ba}{\begin{array}}
\newcommand{\ea}{\end{array}}

\newcommand{\cL}{{\cal L}}

\newcommand{\annp}[3]{{\it  Ann. Phys. (N.Y.) }{{\bf #1} {(#2)} {#3}}}

\newcommand{\fp}[3]{{\it Fortschr. Phys. } {{\bf #1} {(#2)} {#3}}}

\newcommand{\np}[3]{{\it  Nucl. Phys. }{{\bf #1} {(#2)} {#3}}}
\newcommand{\sovphjetp}[3]{{\it Sov. Phys. JETP }{{\bf #1} {(#2)} {#3}}}
\newcommand{\sovphjetpl}[3]{{\it Sov. Phys. JETP Lett.}{{\bf #1} {(#2)} {#3}}}
\newcommand{\sovjnp}[3]{{\it Sov. J. Nucl. Phys. }{{\bf #1} {(#2)} {#3}}}

\preprint{FIAN/TD/98-17}
\begin{document}

\draft

\title{Collective Excitations in Thermal QED$_{3+1}$:   %$
Survival of the Fittest
}

\author{I.~V.~Tyutin\thanks{e-mail: tyutin@td.lpi.ac.ru} and
Vadim Zeitlin\thanks{e-mail: zeitlin@td.lpi.ac.ru}}

\address{
I.~E.~Tamm Theory Department,  P.~N.~Lebedev Physical Institute
Russia,\\ 117924, Moscow, Leninsky prospect, 53}

%\date{}

\maketitle

\begin{abstract}
The spectrum of collective fermionic excitations in a finite
temperature QED$_{3+1}$      %$
is studied in different  regimes.
It is shown that within the standard perturbation approach the
one-loop dispersion equation, besides the ordinary one-particle
excitation, has four new solutions. The additional excitations are
gauge-dependent and two of them have nonphysical signs of residues in the
propagator poles. The temperature evolution of the
solutions is investigated and it is shown that the use of effective
propagators leaves no more than one additional mode which becomes propagating
at $T$\hbox{\bolpor}$10M$, when the gauge invariance is restored. The other
three modes, including those with nonphysical residues in the propagator
poles, are always strongly damped, thus the thermal effects do not produce
pathologies in  QED$_{3+1}$.   %$

\vspace{0.3cm}

\noindent
PACS number(s): 03.70.+k, 11.10.Wx, 52.60.+h.

\end{abstract}

\section{Introduction}

A significant advacement in studying gauge theories at high
temperatures has been achieved in recent years (see e.g. \refc{S}). Hard
thermal loops (HTL) resummations technique \cite{HTL} has made possible to
penetrate into the temperature region, where the effective coupling constant
is large, $gT/M \gg 1$ ($M$ is the vacuum fermion mass), and the ordinary
perturbative series fail. The low--temperature limit, where the expansion in
the vacuum coupling  constant is
valid, is also well-known (see, e.g. \refc{F}).  However, the intermediate
range of temperatures comes as surprise:  the naive one-loop calculation in
the framework of QED shows that at $T$\hbox{\bolpor}$4 M$ new modes arise in
the fermion spectrum.

At low temperature the modification of the fermion dispersion equation is
obvious:  thermal effects produce only a small shift of the vacuum mass shell
of the one-particle excitation (OPE) and no new branches in the spectrum are
formed.  At high temperature the loop corrections may be of a tree-level
order. For instance, when calculating the fermion self-energy every
perturbation order brings the factor $T^2$ (which is typical for diagrams
containing hard thermal loops, electron self-energy among them) and the
effective coupling constant $gT/M$ becomes (unlimitedly) large and one must
sum out HTL. In practice, HTL summation implies substitution of the
effective propagators, which are functions of the effective masses and
dampings instead of the bare ones. In (extremely) high temperature limits,
when the vacuum fermion mass is negligible, an additional hole-like state
with a negative frequency arises in the electron branch, as well as a state
with a positive frequency in the positron branch. Thus the number of
excitations doubles as compared to the low--temperature case.  The additional
solutions (collective excitations, CE) were found in \refs{Kl,W89,P89},
where the one-loop self-energy of (massless) fermion was calculated (the
spectrum structure at high temperatures has been studied also in
\refs{Kal84,Kal97}).  Masses of the above excitations are of the order of
$gT$, and in the leading approximations the corresponding solutions to the
dispersion equation are gauge-invariant \cite{HTL,Kl}.

Early in the 90th it was shown in \refc{LS} in the high--temperature limit
and in \refc{BBS} at the "naive" one-loop level in QED with massive fermions
that at $T > 4M$ in each branch a new low-frequency mode arises.
Besides, it was demonstrated in \refc{LS} that the above excitation
disappears after the damping of OPE is taken into account in the effective
propagator.

Therefore, several questions arise: how high must the temperature be to let
collective excitations come into existence, how do they exactly evolve
with temperature, and are they physical excitations or not?  In the present
paper we shall study properties of the additional solutions to the dispersion
equation, point out the physical ones, and estimate the temperature required
to make them propagating.  Within the naive perturbation theory the
additional modes arise at  rather low temperatures, $T$\hbox{\bolpor}$4M$.
These modes are situated in the low frequency range ($\omega\sim M^3/T^2$ at
zero momentum) and always arise in couples:   two modes at positive and two
at negative frequencies, both on electron and positron branches. However, the
additional solutions are pathological:  one excitation in each pair has the
"wrong" sign of the residue. Besides, all the modes are (in leading order)
gauge dependent. When the temperature is raised the additional
positive-frequency modes (in the electron branch)  and one of the
negative-frequency modes (that with the wrong residue sign) remain in the
low-frequency region, while the absolute value of frequency of the other
modes rise and becomes $\sim gT$.

Since the difficulties described above jeopardize not an
abstract model, but quantum electrodynamics, the theory describing the
real world, one should believe the pathological modes (those which are
gauge-variant and/or have indefinite metrics) should be exorcised by applying
a correct calculation procedure.

We shall demonstrate that after inserting the damping of OPE into the
bare propagators among positive frequency exitatons only one mode survives at
all temperatures.  At  $T$\hbox{\bolpor}$5M$ the real part of the
inverse propagator has two zeroes at negative frequencies, but its
imaginary part is larger than the frequency, thus the
physical excitations are absent, and only at $T$\hbox{\bolpor}$10M$ one of
the negative-frequency solutions overcomes the damping and becomes the
propagating one. This is in a good agreement with the approach to the region
$T\sim 10M$ from the higher temperature domain where the hard thermal
loops results are valid.

The paper is organized as follows. In Sec. 2 the one--loop self--energy in an
arbitrary relativistic gauge is calculated using the bare propagators. We
show that the above naive calculation brings pathological modes into the
model. In Sec. 3 we study how the damping may affect the collective
excitations in QED plasma and show that the "improved" naive approach and HTL
scheme are in a good agreement with one another in the intermediate
temperature range.  A possible physical interpretation of the additional
high-temperature excitation found in the fermion spectrum is given in Sec. 4.

\section{One--loop electron self--energy at finite temperature}

In this section we shall study the spectrum of one--particle fermion
excitations within the standard one-loop approximation, i.e. using the free
electron and photon propagators. We shall use the thermal Green functions
method \cite{F,M,FJ,AGD} (of course, the real-time approach \cite{K,LL,S}
produces the same results).
The spectrum is defined as poles of the Fourier--transformed retarded Green
function  $S_R(p_0,{\bf p})=S_R(p)$ or, equivalently, as zeroes of
$S^{-1}_R(p)$.  The propagator $S_R(p_0,{\bf p})$ is evaluated from
the thermal Green function $S(i\omega_n,{\bf p})$, $\omega_n=(2n+1)\pi\b$,
$\b\equiv 1/T$ via analytical continuation $i\omega_n$ $\rightarrow$
$\omega\equiv p_0$, holomorphic in the upper half-plane of $\omega$ and having
correct asymptotic behavior at $\omega$ $\rightarrow$ $\infty$ (the fermion
self-energy falls as  $1/\omega$ at high frequencies) \cite{F,AGD}.

The Lagrangian of QED$_{3+1}$ in a relativistic gauge  $\alpha$ is

        \be
        \cL = -\frac14 F_{\mu\nu}F^{\mu\nu}
        - \frac1{2\alpha} (\d_\mu A^\mu)^2+
        \bar{\psi}
        (\imath \hat{\partial} + g \hat{A} -M)\psi~~~.
        \ee

The fermion self-energy  $\Sigma(p)$ is defined as \footnote{We
define the Green functions (propagators) in the temperature technique as
average of the $T$--products in thermal time $\tau$ of field operators.}:

        \be
        S^{-1}(p)=S^{-1}_0(p)+\Sigma(p), \quad p_0=i\omega_n~~~,
        \ee

        \be
        S_0(p)=-\frac1{\hp-M}~~~.
        \ee

\noi
In the one--loop approximation it equal to

        \be
        \Sigma_1(p)=- \i g^2T  \sum_l\int \frac{d {\bf k}}{(2\pi)^3} \g_\mu
        S_0(p+k) \g_\nu D^{\mu\nu}(k)~~~,~~~k_0=i{2\pi l\over T}~~~,
        \label{s1}
        \ee

       \be
        D_{\mu\nu}(k) =\frac{1}{k^2}
        \left(
                g_{\mu\nu} - \frac{k_\mu k_\nu}{k^2} \right)
                        +\alpha \frac{k_\mu k_\nu}{k^4}~~~.
        \label{dmunu}
        \ee

The study of $\Sigma_1(p)$ for arbitrary temperature, frequency and external
3--momentum ${\bf p}$ is quite difficult and we consider the case of zero
spatial momentum, ${\bf p=0}$, only.  The corresponding expression for
$\Sigma_1$ may be written as (in  Ref. \cite{BBS} $\Sigma_1(\o)$ was
calculated in the Feynman gauge):

        \be
        \Sigma_1(\omega,{\bf 0})\equiv\Sigma_1(\omega)=\gamma^0
        \left(a(\omega)+(1-\alpha)b(\omega)\right)+
        {3+\alpha\over4}c(\omega)~~~,
        \label{sigma}
        \ee

        \be
        \gamma^0=\left( \matrix {1 &0 \cr 0 & -1 \cr}
 %       \begin{array}{rr}
 %       1&0\\0&-1
 %       \end{array}
        \right)~~~,
        \ee

        \be
        a=\frac{g^2\o}{\pi^2}
        \int_0^\infty \frac{d\! k}{4\o^2k^2-(M^2-\o^2)^2}
        \left(
        \frac{k(M^2-\o^2+2k^2)}{e^{\b k}-1}
        +
        \frac{2k^2\ve}{e^{\b \ve}+1}                    \right)~~~,
        \label{a}
        \ee

        \bea
        \lefteqn{b=\frac{ g^2\o (M^2-\o^2)}{2\pi^2}
        \int_0^\infty \frac{d\! k}{(4\o^2k^2-(M^2-\o^2)^2)^2}\times}\\
        \label{b}
        &&\nonumber\\
        &&\left(
        \frac{k(8\o^2k^2 -4k^2M^2-(M^2-\o^2)^2)}{e^{\b k}-1}
        +
        \frac{k^2((M^2+\o^2)^2-4\ve^2M^2)} {e^{\b\ve}+1}
                                                \right)~~~,\nonumber
        \eea

        \be
        c=\frac{2 g^2 M}{\pi^2}
        \int_0^\infty \frac{d\! k}{4\o^2k^2-(M^2-\o^2)^2}
        \left(
        \frac{k(M^2-\o^2)}{e^{\b k}-1}
        +
        \frac{k^2(M^2+\o^2)}{\ve(e^{\b \ve}+1)}
        \right)~~~
        \label{c}
        \ee

\noi
( the vacuum contribution, which is insufficient at high temperatures, is
omitted).

The spectrum is defined as nontrivial solutions to the equation

        \be
        (\gamma^0\omega-M - \Sigma_1(\omega))\psi=0~~~.
        \label{ud}
        \ee

We suppose that an excitation belongs to the electron branch if $\psi$
may be written as

        \be
        \psi=\left(
%\begin{array}{c}u\\0\end{array}
 \matrix{u\cr 0\cr}
\right)~~~,
        \ee

\noi
and to the positron branch if  $\psi$ is

        \be
        \psi=\left(
%\begin{array}{c}0\\v\end{array}
 \matrix{0\cr v \cr}
\right)~~~.
        \ee

\noi
We shall consider below the electron branch excitations only, since the
positron branch solutions may be obtained by reversing the sign of
$\omega$. The electron spectrum is determined as solutions to the following
equation:

        \be
        \omega-M=a(\omega)+(1-\alpha)b(\omega)+\dfrac{3+\a}4 c(\omega)~~~.
        \ee

In the vicinity of a pole the retarded propagator may be presented as

        \be
        S_R(\omega)\sim Z_E{\psi_E\bar{\psi}_E\over \omega-E+i\delta}~~~,
        \ee

\noi
$\omega=E$ is a solution to the dispersion equation, $\psi_E$ is the
 solution to the \Eq{ud} for $\omega=E$ normalized to unity, and the residue
$Z_E$ is

        \be
        {1\over Z_E}=\left.{\partial\over\partial\omega}(\omega-M-
        \Sigma_1(\omega))\right|_{\omega=E}~~~.
        \ee

At low temperature corrections to the free propagator are small and the
dispersion equation has just one solution, $\omega \approx M$.
When the temperature is raised, $\Sigma_1(\o)$ increases as $T^2$, but for
low frequencies $\Sigma_1(\o)$ varies drastically due to the
contribution to the integral of the momentum region corresponding to
pole in the integrand (this happens since the poles in the photon and
electron propagators in the expression for $\Sigma_1$, \Eq{s1} are very
close to one another).

In Fig. 1 the dependence of real part of the fermion self-energy on frequency
for the electron branch in the Feynman gauge  $\alpha=1$ (i.e.  the function
$a(\omega)+c(\omega)$) is presented for $T=3M$ and $5M$.  The intersections of
$\Re e \Sigma_1$ with the straight line  $\omega-M$ are solutions to
the dispersion equation in the electron sector. One can see that starting
with $T$\hbox{\bolpor}$4M$ there are five solutions in each branch.
It follows from the continuity of $\Sigma(\omega)$ as a
function of $\o$ that the additional solutions come by two (the plot of
$\Re e \Sigma_1$ at $T=5M$ in Fig.1 confirms this) and that the residues of
two poles in each pair have opposite signs.  Actually, the exact location
and the residue of the poles depend on the gauge.
This follows from the explicit form of the function $\Sigma_1(\omega)$
\Eq{sigma}. Consider the pole closest to zero as an illustration. Its
location and residue may be obtained by expanding $\Sigma_1(\omega)$ in
series in $\omega$ (the coefficients are calculated in the leading order in
$T/M$):

        \be
        \omega=\dfrac{ M+{3+\alpha\over4}c}{1+(2-\alpha)q}~~~, \quad
        Z=1+(2-\alpha)q~~~,\label{low_o}
        \ee

\noi
with

        \be
        q={2g^2\over\pi^2}{T^4\over M^4}\int_0^\infty k^3 d\! k\left(
        \frac{1}{e^k-1}+\frac{1}{e^k+1}\right)\approx-a'(0)\approx-b'(0)~~~,
        \label{low_a}
        \ee

        \be
        c={2g^2\over\pi^2}{T^2\over M}\int_0^\infty k d\! k\left(
        \frac{1}{e^k-1}+\frac{1}{e^k+1}\right)\approx c(0)~~~,
        \label{low_c}
        \ee

\noi
i.e. parameters of this excitation do depend on $\a$.

Thus we see that the additional excitations arising in the perturbation
theory are pathological: their characteristics are gauge--variant and some
residues are negative, which implies indefinite metrics. One should believe
that when a correct calculational procedure is used the excitations with
unacceptable properties would vanish from the physical spectrum.

When the temperature increases excitations behave in a differentl way.
Energies of the middle three remain small. This may be easily understood:
the above excitations arise due to large oscillations of $\Sigma_1$  in the
small frequency region. As we have discussed above, those oscillations are
due to large contribution of the momenta corresponding to the pole of the
integrand in $\Sigma_1$.  It is well-known that at high temperatures the
effective mass of OPE becomes much greater than the vacuum one. If one
neglects the vacuum mass while calculating $\Sigma_1$, the dispersion
equation has two solutions for the electron branch (and two for the positron
one), then three middle excitations vanish, and the energies of the remaining
modes grow when the temperature is raised and the corresponding poles prove
to be gauge-invariant in the leading order in temperature \cite{HTL,Kl}.

Therefore, the following picture may be expected: within a correct
calculational procedure the pole in the integrand of  $\Sigma$ would be
somehow regularized, and the regularization would affect the low frequency
region only. At low and medium energies the dispersion equation would have
only one solution for the electron and only one for the positron branch. When
the temperature grows an additional solution will appear, the same as in the
high temperature limit.

\section{Account taken of damping}

When calculating the Feynman diagram in quantum field theory or in the
real-time approach to the quantum statistics a correct definition
(regularization) of propagator poles is necessary. This is achieved
by adding infinitesimal imaginary term  $i\delta$ to $p_0$ or $p^2$
depending on the propagator type, that makes the propagator free of
singularities when $\delta\neq0$. Thus a natural method of regularization is
the use of propagators with nonzero imaginary part (this approach to
regularization of the integrand in  $\Sigma$ was proposed in \cite{LS}).  The
above propagator arises in the following way. The exact expression for the
fermion self-energy is

        \be
        \Sigma(p) = g^2 \int \frac{d^4\! k}{(2\pi)^4}\gamma^\mu
        S(p-k)\Gamma^\nu(p-k,p)D_{\mu\nu}(k)~~~,
        \ee

\noi
where $S$, $D_{\mu\nu}$, $\Gamma^\nu$ are the exact fermion and photon
propagators and the vertex (within the temperature technique the
4--dimensional integral $\int d^4k$ must be replaced by $-2\pi T\sum_n\int
d^3k$). At high temperature, $M\ll|\omega|\ll T$, where HTL
approach is valid, the free vertex  $\gamma^\nu$ is used as $\Gamma^\nu$, and
propagators are free {\it effective} propagators, which are functions of
effective masses and fermion damping (see below). At low temperature the
self-energy must be calculated using the standard perturbation expansion.
However, we shall replace the free fermion propagator by free {\it damped}
fermion propagator, i.e.  that containing OPE damping  $\gamma$.  For the
retarded propagator it is equivalent to the substitution $p_0 \to
p_0+i\gamma$, thus in the temperature technique we have the standard equation
with $\omega_n$ replaced by $\omega_n+\gamma$. For $\gamma$ at low
temperature we shall take the high temperature expression \cite{LS},

        \be
        \gamma= -\frac{g^2T}{12}\ln g~~~.
        \label{zatuhanie}
        \ee

As a result, for $\Sigma(\omega)$ at low temperatures one obtains
equations (6) -- (9) again, where the substitution $\o \to \o+i\gamma$ is
made (we have neglected the imaginary term $i\gamma$ in the thermal
distribution functions, since the principal contribution to the integral
comes from the momenta $k\sim T$, and  $\gamma\ll k$).

In Fig. 2 plots of imaginary and real parts of the electron branch of the
self-energy calculated in the Feynman gauge
at $T=5M$, $8M$ and $10M$ are presented. One can notice that there are no
additional solutions at positive frequency. At negative
frequencies the real part of the inverse propagator has two zeroes, which
would imply existence of two additional solutions. However, the right
(closest to $\omega=0$) solution to the equation  $\Re e (\o - M - \Sigma)=0$
is always within the range of large (negative) imaginary part of
$\Sigma(\omega)$, $\Re e E\ll \Im mE$, and thus the physical excitations are
absent.  The left zero at $T=5M$ and $T=8M$ (Fig.  2 a, b) is also strongly
damped. However, when the temperature is increased the corresponding point
moves leftward and at $T$\hbox{\bolpor}$10M$ escapes the high damping domain
(Fig.  2 c).  Therefore, the temperature for which an additional
(propagating) mode appears is approximately $10M$.

At  $T\gg |\omega| \gg M$ the fermion self-energy $\Sigma$ should be
calculated using the effective electron propagator $S_{eff}$

        \be
        S_{eff}(p)=-{\gamma^\mu
        p_\mu\over p_0^2-p_i^2-M^2_{eff}},\quad p_0=\i {(2n+1)\pi\over
        T}+\i \gamma,\quad M^2_{eff} = \frac{g^2T^2}4
        \ee

\noi
and the photon propagator \footnote{ In \refc{LS} the photon propagator
was taken in the $A_0=0$ gauge.  We are using the Feynman gauge, since the
results are leading order gauge-invariant.} $D^{eff}_{\mu\nu}(k)$

        \be
         D^{eff}_{\mu\nu}(k)={g_{\mu\nu}\over
        k^2-\mu_{eff}^2}~~~,\quad k_0=i{2n\pi\over T}~~~, \quad
        \mu^2_{eff}=\frac{g^2T^2}{12}~~~.
        \ee

In the above temperature range the self-energy in the leading order in $T$ is

        \be
        \Sigma(\omega)= {M^2_{eff}\over2 \o}~~~
        \ee

\noi
and the dispersion equation looks like

        \be
        \omega = {M^2_{eff}\over2 \o}~~~.
        \label{25}
        \ee

\noi
This provides the following value of OPE mass $\o_+$ (energy at zero
momentum)

        \be
        \o_+ = \dfrac{1}{\sqrt{2}} M_{eff} = \dfrac{gT}{2\sqrt{2}}
        \ee

\noi
and CE mass $\o_-$

        \be
        \o_- = -\dfrac{1}{\sqrt{2}} M_{eff} = -\dfrac{gT}{2\sqrt{2}}~~~.
        \ee

To evaluate the localization of the additional low--frequency excitation
localization one may use Eq.  (17) with the substitution  $M \to M_{eff} \sim
gT$, which gives

        \be
        \frac{|\o|}{M_{eff}}
        \,\hbox{\menpor} g^2 ~~~,~~~|\omega|\,\hbox{\menpor} g^3T
        \ee

\noi
The plot of $\Im m\Sigma(\omega)$  presented in Fig. 3 demonstrates that the
imaginary part of $\Sigma$ is large in this frequency domain (the fact that
insertion of OPE damping into the effective propagator makes $\Im m\Sigma(0)
\sim T$ was mentioned in \refc{LS}).

Let us check whether the results for CE energy at $T=10M$  where the
effective HTL scheme (a descent from high temperatures to $T=10M$) is
consistent with that obtained within the "improved" naive approach (an ascent
from low temperatures to $T=10M$). In Fig. 4 the plot of  $\Re
e\Sigma(\omega)/M_{eff}$ (the quantity independent of temperature) calculated
within the HTL method, and the straight line $(\o-M)/M_{eff}$ for $T= 10$ are
presented. (One cannot yet neglect the vacuum mass in the dispersion equation
at this temperature since $M_{eff}$ and $M$ are of same order.)  It
follows from Fig. 2c and Fig. 4 that the values of CE (and OPE) energies are
perfectly matched.

Let us emphasize that although the exact propagator has two poles one should
take into account only one (OPE pole) when calculating $\Sigma$, since the
residue in CE pole at $k\sim T$ is exponentially small \cite{W89}.

\section{Physical interpretation}

A satisfactory mechanism giving rise to an additional
collective excitation is not revealed yet. In \refc{W89} the appearance
of CE was supposed to be a result of interaction between  the electron and
a sea of electron-positron pairs. We shall show below that in the model
describing "nonrelativistic" electron and photon, when antiparticles are
absent, a CE similar to that in quantum electrodynamics still arises.

Consider the Lagrangian

        \be
        L=\psi^\dagger(\i\partial_t-\varepsilon(\i{\vec {\partial}}))\psi+
        {1\over2}(\partial_t\varphi\partial_t\varphi-
        \varphi\kappa^2(\i{\vec{\d}})\varphi)+
        \psi^\dagger\psi f(\i{\vec {\d}})\varphi~~~.
        \label{l2}
        \ee

\noi
The self-energy $\Sigma(p)$ of the field $\psi$ is defined as

        \be
        S^{-1}(p)=S^{-1}_0(p)+\Sigma(p)~~~,
        \ee

\noi
where $S$ and $S_0$ are the exact and the free propagators of the field $\psi$.
At the one--loop level the self--energy is

        \be
        \Sigma_1(p)=\int{d^3k\over(2\pi)^3}{f^2(k)\over
        \kappa^2(k) -(p_0-\varepsilon(p-k))^2}\left[
        {\varepsilon(p-k)-p_0\over\kappa(k)}
        {1\over e^{{\b\kappa(k)}}-1} +
        {1\over e^{{\b\varepsilon(p-k)}}+1}             \right]
        \ee

\noi
(the vacuum contribution is omitted).

First, suppose that

        \be
        f(k)\equiv g,\quad \varepsilon(k)=\sqrt{k^2+M^2}~~~,\quad \kappa(k)=
        \sqrt{k^2+\mu^2}~~~.
        \ee

\noi
In this case \Eq{l2} describes a model with relativistic kinematics like
electrodynamics, but no antifermions.
Let the external momentum  ${\bf p}$ be zero and consider the frequency range
$M,\mu\ll|\omega|\ll T$, $\omega\equiv p_0$.  Taking into account
the fact that the principal contribution comes from the integration
over $k\sim T$  one obtains in the leading order in $T$:

        \be
        \Sigma_1(\omega)={1\over\omega}{g^2T^2\over4\pi^2}
        \int_0^\infty xdx
        \left[{1\over e^x-1}+{1\over e^x+1}\right]=\dfrac{g^2T^2}{16\o}~~~.
        \label{sotr}
        \ee

The corresponding dispersion equation may be written as

        \be
        \omega={\omega_0^2\over\omega}, \quad \omega^2_0=
        \dfrac{g^2T^2}{16}~~~,
        \ee

\noi
which coincides with the similar equation in QED, \Eq{25}.

For negative frequencies the self-energy is a negative quantity
(at $M>\mu$), in particular,

        \be
        \Sigma_1(0)=\dfrac{g^2}{2\pi^2(\mu^2-M^2)}
        \int_0^\infty k^2dk
        \left[
        {\varepsilon(k)\over\kappa(k)}
        {1\over e^{{\b\kappa(k)}}-1}+
        {1\over e^{{\b\varepsilon(k)}}+1}             \right]
        \approx \dfrac{7 g^2 T^3 \zeta(3)}{4\pi^2(\mu^2-M^2)}
        \ee

\noi
(the approximate equality is valid for $T\gg M,\mu$).  Since $\Sigma_1(\o)$
is positive at high positive frequencies (see \Eq{sotr}) and  $|\Sigma_1(0)|$
grows and becomes greater than $M$ when the temperature increases, the
situation is analogous to QED: at low temperature  $\Sigma_1(\o) $  is small
and only one solution (corresponding to OPE) exists. When the temperature is
growing two additional solutions appear, one of them having negative residue
in the propagator pole. When the temperature raised further on, one of the
additional solutions is shifting to large negative frequencies (and it would
leave the damping domain if we take the OPE damping into account when
calculating $\Sigma$). The frequency of the second solution remains small and
has  negative residue (the imaginary part of $\Sigma$ would be large there
if we took the damping into account).

Second, consider the electron--phonon interaction model

        \be
        f^2(k)=g^2\kappa^2(k)\theta(k_D-k)~~~, \quad g^2={2\pi^2\over
        k_DM}~~~,\quad \varepsilon(k)={k^2\over2M}-\mu~~~,\quad
        \kappa(k)=ck~~~, \ee

\noi
where $M$ is the electron mass, $c$ is the speed of sound, $k_D\sim p_F$,
$\mu\sim p^2_F/2M$, $p_F$ is Fermi momentum. Let the external momentum
be $p=p_F$ and let us examine the frequency range $|\omega|\gg p^2_F/M$. In
this case in the leading order in $T$ one has

        \be
        \Sigma_1(\omega)={p^2_FT\over M}{1\over\omega}
        \ee

\noi
and the corresponding dispersion equation is

       \be
        \omega={\omega_0^2\over\omega},\quad \omega^2_0={p^2_F\over M}T~~~.
        \ee

Therefore, starting with $T$\hbox{\bolpor}$p^2_F/M$ the exact
propagator of the field $\psi$ has poles in symmetric points:

        \be
        \omega=\pm\omega_0=\pm\sqrt{{p^2_F\over M}T}~~~.
        \ee

A typical Fermi energy is few $eV$, thus a pole
in the symmetric point $\omega=-\omega_0$ would appear at
$T$\hbox{\bolpor}$10^4$ -- $10^5K$.  At these temperatures the model
considered is apparently senseless, moreover even "solids" do not
exist.  But we come to an important conclusion: just in any model, not only
relativistic ones, at sufficiently high temperatures a pole in a symmetric
point $\omega=-\omega_0$ arises. That symmetric pole is a result of the
growth of $\Sigma$ and looks like an increase of the effective
coupling constant. The latter may be explained as follows: when the
temperature is raised the characteristic momentum grows, $k\sim T$, and the
number of particles participating in the interaction (phase volume) increases
drastically.

\bigskip
Authors are grateful to P.I.Arseev, O.K.Kalashnikov, V.V.Losyakov,
E.G.Maksimov and A.V.Smilga for fruitful discussions and to A.E.Shabad for
cooperation and critical comments on the manuscript.  This work was
supported in part by grants RFBR 96-02-17314 and INTAS-RFBR 95-829 (I.~T.)
and  RFBR 96-02-16117 RFBR 96-02-16210 (V.~Z.).

\newpage

\begin{figure}

\centerline{\psfig{figure=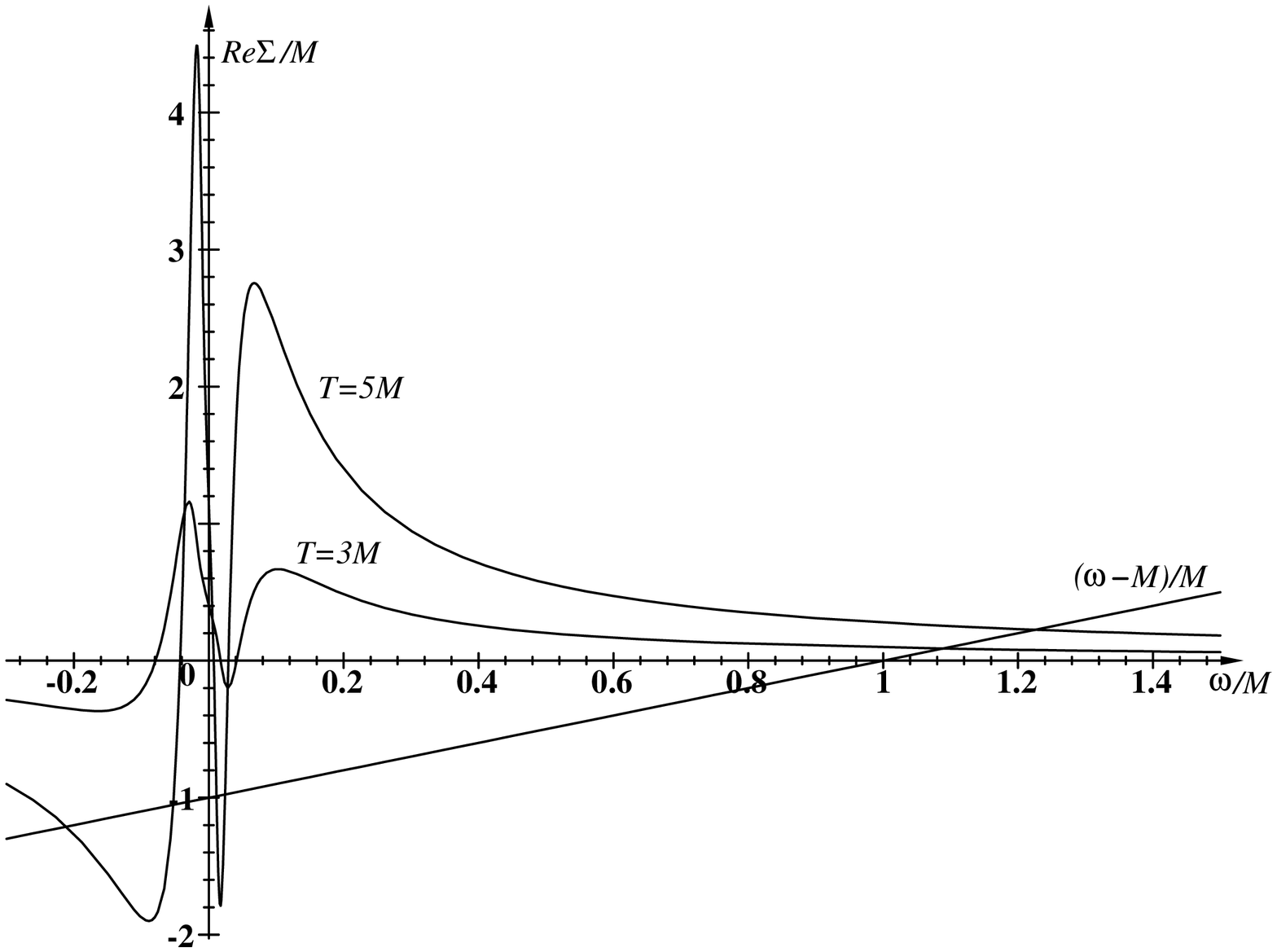,width=6.5in,angle=-90}}
\caption{Real part of one--loop fermion self-energy at $T=3M$ and $T=5M$
and $\omega - M$ (straight line).  Frequency and
self-energy are normalized by fermion vacuum mass.  } \label{fig1}
\end{figure}

\newpage
\begin{figure}

\phantom{a}

\vspace{-3.cm}
\centerline{\psfig{figure=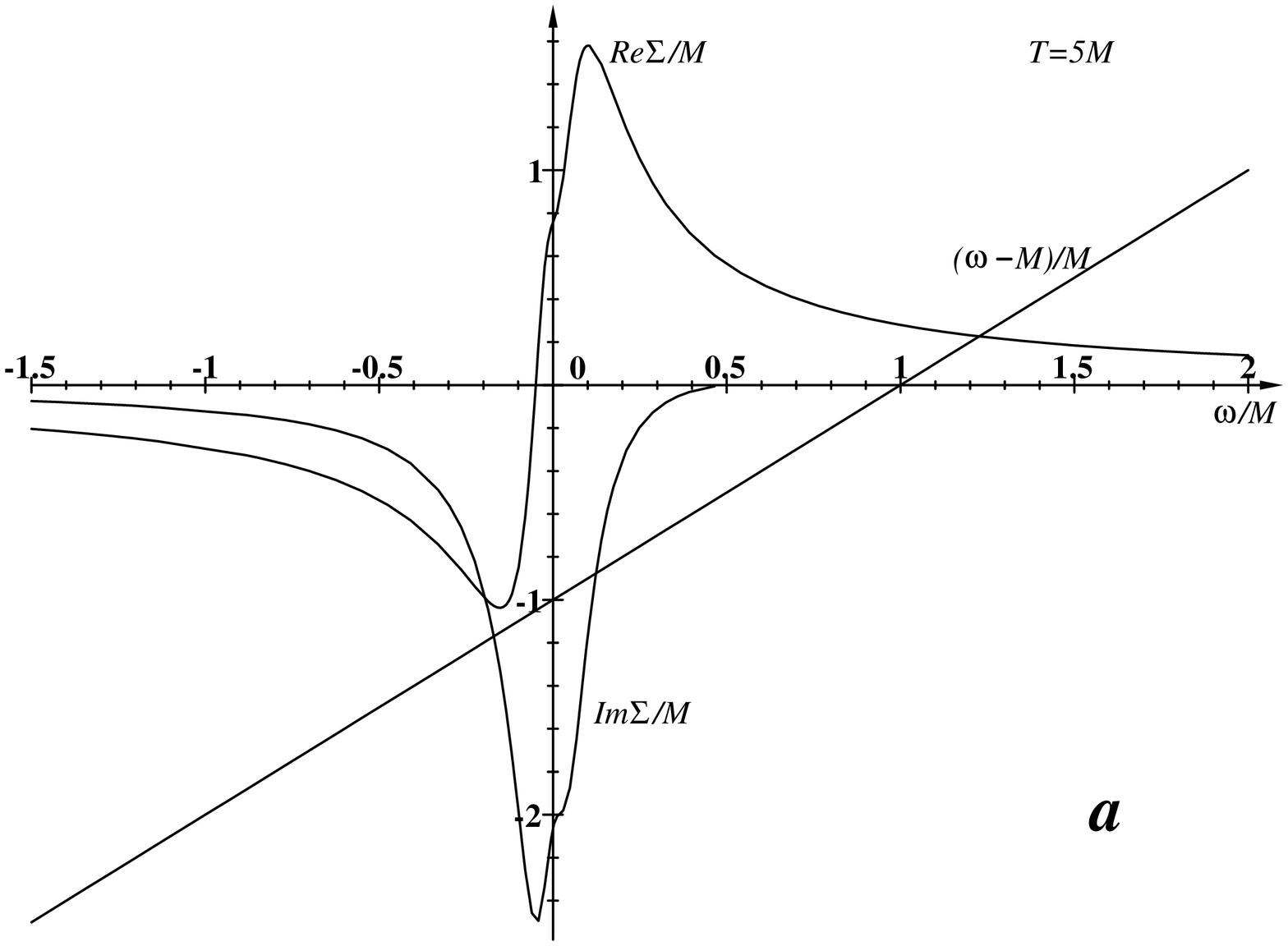,width=4.8in,angle=-90}}

\vspace{-2.2cm}
\centerline{\psfig{figure=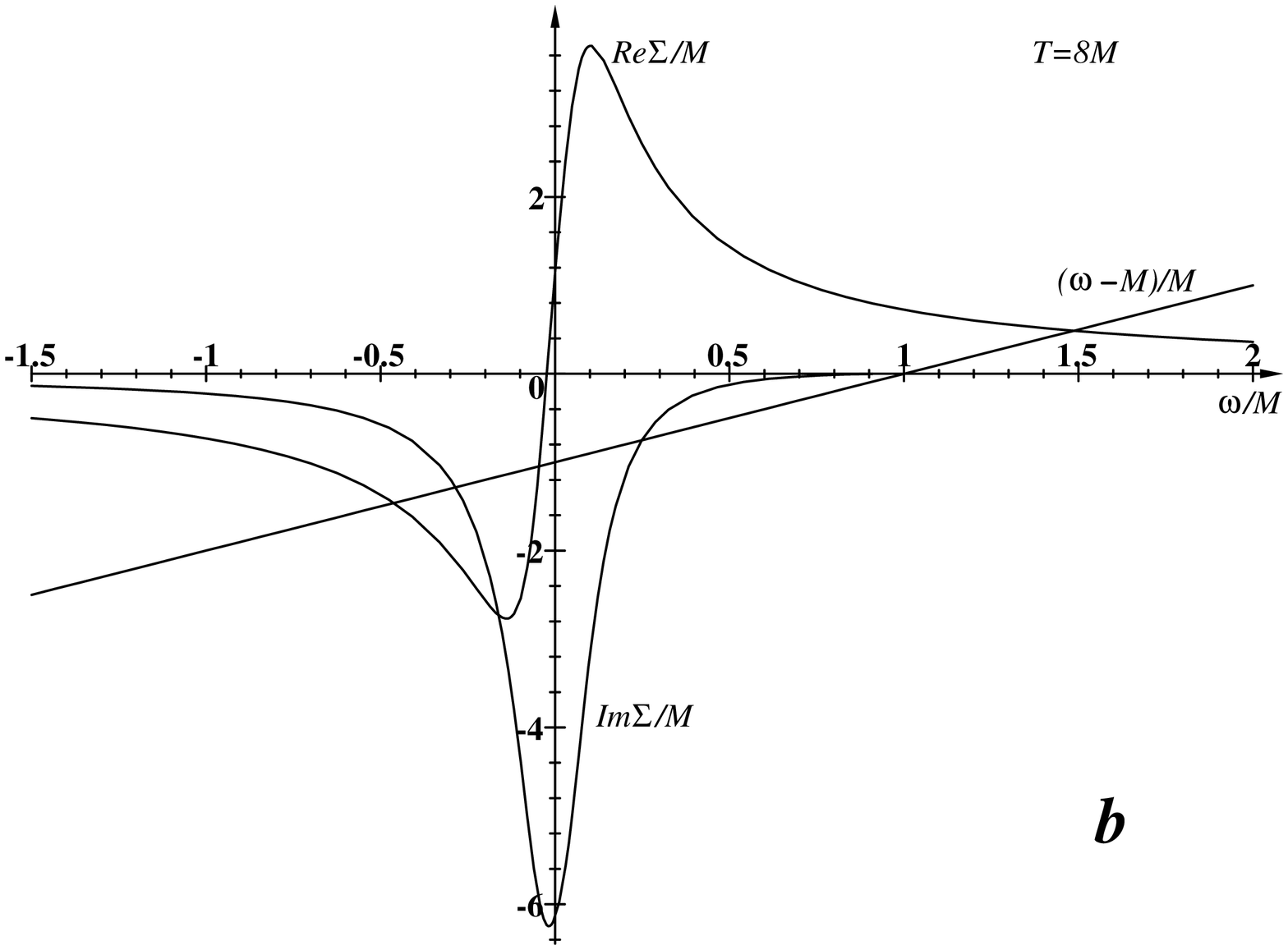,width=4.8in,angle=-90}}

\vspace{-2.2cm}
\centerline{\psfig{figure=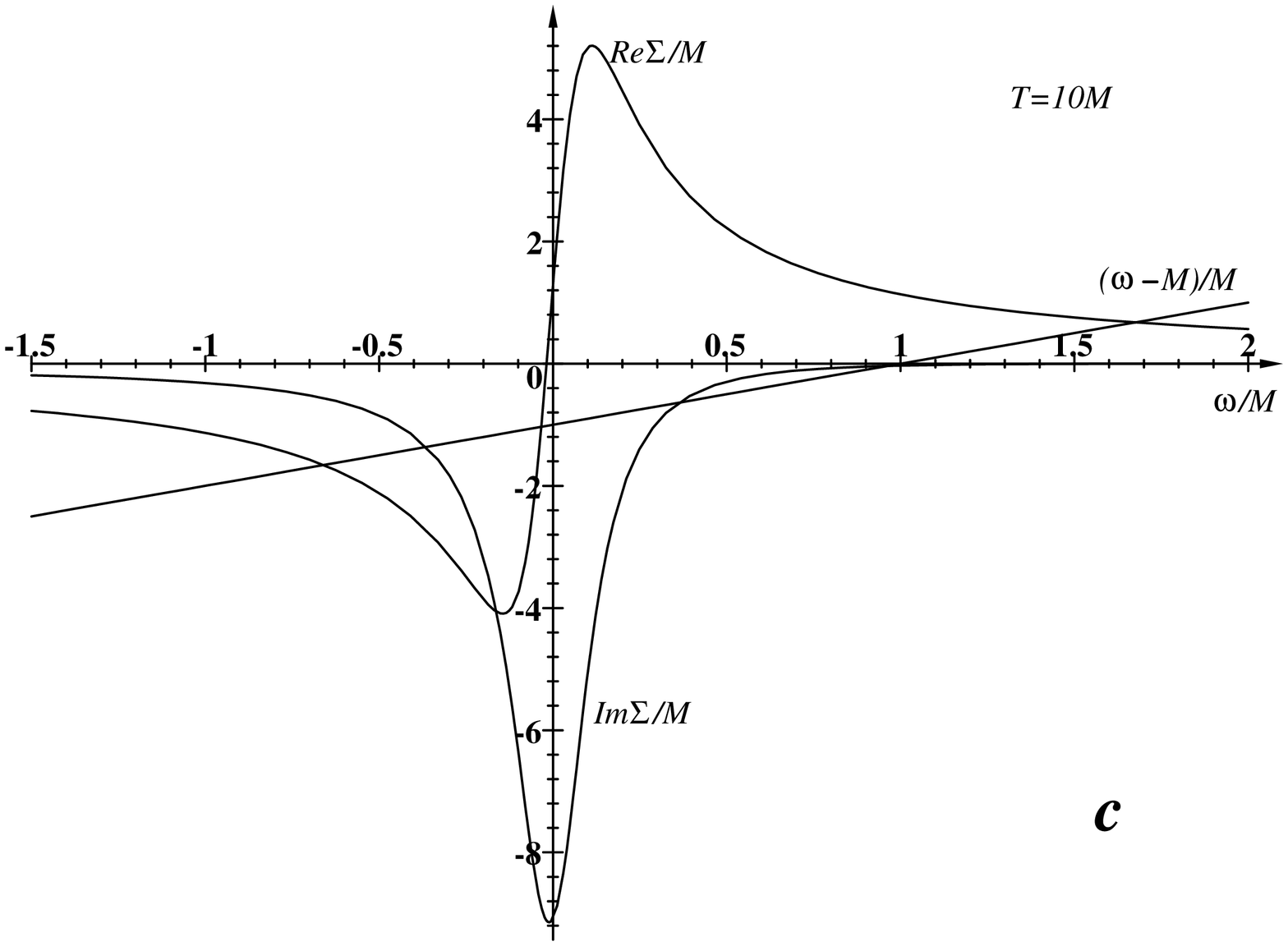,width=4.8in,angle=-90}}
\caption{Imaginary and real parts of $\Sigma$ at $T=5M, 8M$ and $10M$
(frequency and self-energy are normalized by fermion vacuum mass).}
\label{fig2}
\end{figure}

\newpage
\begin{figure}

\centerline{\psfig{figure=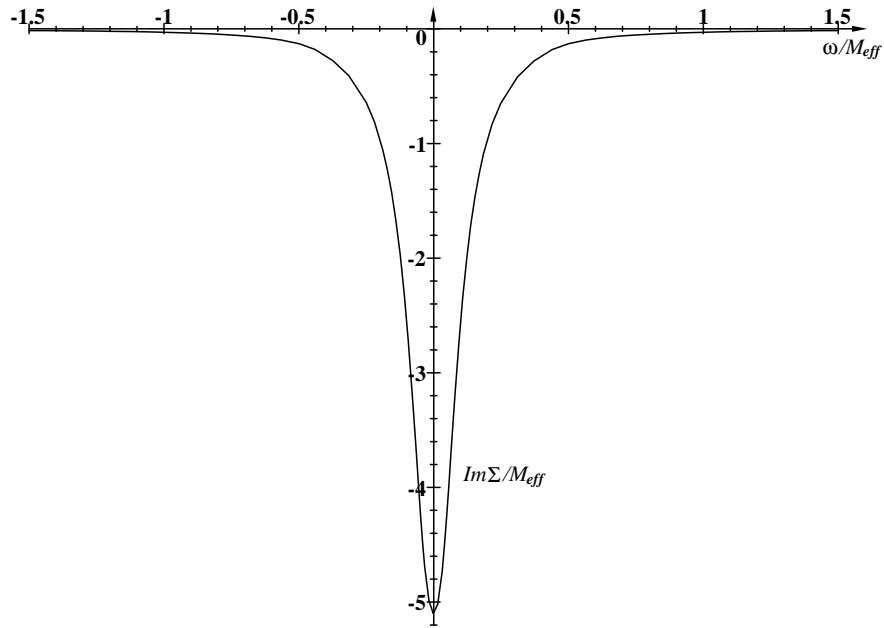,width=4.5in,angle=-90}}
\caption{$\Im m\Sigma$ as function of frequency
($\o$ and $\Sigma$ are normalized by $M_{eff}=gT/2$). Within low frequency
range  $|\o|/M_{eff}< g^2 \sim 0.01$, $\Im m\Sigma/M_{eff}$ is large,
$|\Im m\Sigma|/M_{eff} \sim 5$.}
\label{fig3} \end{figure}

\begin{figure}

\centerline{\psfig{figure=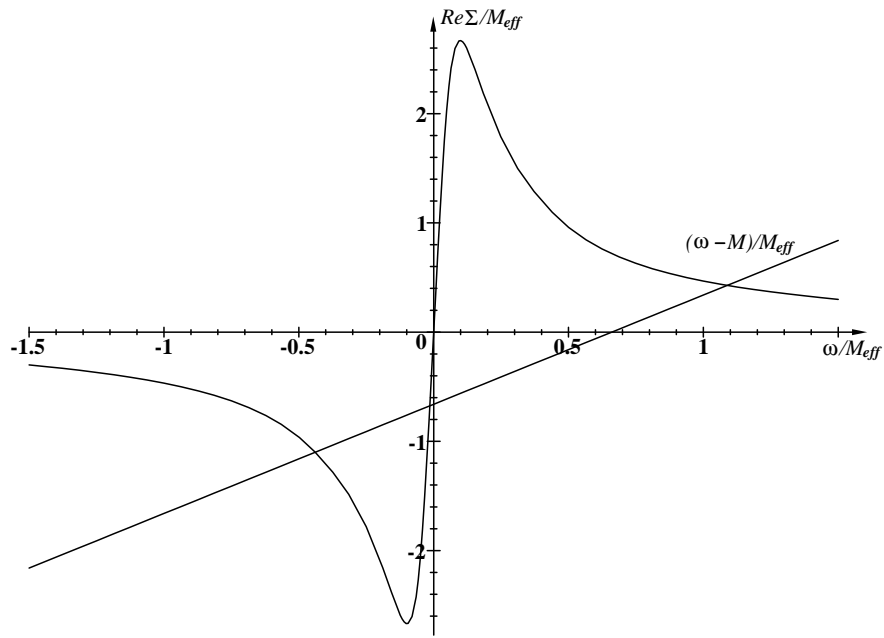,width=4.5in,angle=-90}}
\caption{$\Re e\Sigma$ as function of frequency ($\o$ and $\Sigma$ are
normalized by $M_{eff}$). Left solution $|\o_-|/M_{eff}\approx 0.43$ or
$|\o_-|= 0.64M$.  The latter should be compared with corresponding solution
on Fig. 2c $|\o|_-\approx 0.66M$.}
\label{fig4}
\end{figure}

\end{document}